\documentclass[runningheads]{llncs}

\usepackage{graphicx}
\usepackage{makecell}
\usepackage{subfig} 
\usepackage{wrapfig} 
\usepackage{booktabs}
\usepackage{multirow}
\usepackage{amsfonts}
\usepackage{todonotes}
\usepackage{xcolor}

\def\coo{\small\textsc{CoOccur}}
\def\embedding{\small\textsc{Embedding}}
\def\full{\small\textsc{FullText}}
\def\kwdpubmed{\small\textsc{KwdLarge}}
\def\kwdpmc{\small\textsc{Kwd}}
\def\copmc{\small\textsc{CoOccur-Kwd}}
\def\copubmed{\small\textsc{CoOccur-KwdLarge}}
\def\cofull{\small\textsc{CoOccur-FullText}}
\newcommand{\collectionlong}{\textit{\textbf{D}isease-\textbf{S}ymptom \textbf{R}elation Collection}}
\newcommand{\collectionshort}{\textsc{dsr}-collection}
\newcommand{\collectionshorttitle}{DSR}

\begin{document}

\title{\collectionshorttitle{}: A Collection for the Evaluation \\ of Graded Disease-Symptom Relations}

\author{Markus Zlabinger\inst{1} \and Sebastian Hofst{\"a}tter\inst{1} \and Navid Rekabsaz\inst{2} \and Allan Hanbury\inst{1}}

\authorrunning{M. Zlabinger et al.}

\institute{TU Wien, Vienna, Austria \\ \email{\{\textit{first.last}\}@tuwien.ac.at} \and Johannes Kepler University, Linz, Austria\\~\email{navid.rekabsaz@jku.at}}

\maketitle

\vspace{-20pt}
\begin{abstract}
The effective extraction of ranked disease-symptom relationships is a critical component in various medical tasks, including computer-assisted medical diagnosis or the discovery of unexpected associations between diseases. While existing disease-symptom relationship extraction methods are used as the foundation in the various medical tasks, no collection is available to systematically evaluate the performance of such methods. In this paper, we introduce the \collectionlong{} (\collectionshort{}), created by five physicians as expert annotators. We provide graded symptom judgments for diseases by differentiating between \textit{relevant symptoms} and \textit{primary symptoms}. Further, we provide several strong baselines, based on the methods used in previous studies. The first method is based on word embeddings, and the second on co-occurrences of MeSH-keywords of medical articles. For the co-occurrence method, we propose an adaption in which not only keywords are considered, but also the full text of medical articles. The evaluation on the \collectionshort{} shows the effectiveness of the proposed adaption in terms of nDCG, precision, and recall. 

\end{abstract}

\vspace{-10pt}
\section{Introduction}
\vspace{-10pt}
\label{sec:intro}
Disease-symptom knowledge bases are the foundation for many medical tasks -- including medical diagnosis~\cite{niAutomatedMedicalDiagnosis2017} or the discovery of unexpected associations between diseases~\cite{zhouHumanSymptomsDisease2014,valleEvaluatingWikipediaSource2018}. Most knowledge bases only capture a binary relationship between diseases and symptoms, neglecting the degree of the importance between a symptoms and a disease. For example, abdominal pain and nausea are both symptoms of an appendicitis, but while abdominal pain is a key differentiating factor, nausea does only little to distinguish appendicitis from other diseases of the digestive system. While several disease-symptom extraction methods have been proposed that retrieve a ranked list of symptoms for a disease~\cite{zhouHumanSymptomsDisease2014,shahNeuralNetworksMining2019,martinSymptomExtractionIssue2014,xiaMiningDiseaseSymptomRelation2018}, no collection is available to systematically evaluate the performance of such methods~\cite{shenEnhancingOntologydrivenDiagnostic2019}. While these method are extensively used in downstream tasks, e.g., to increase the accuracy of computer-assisted medical diagnosis~\cite{niAutomatedMedicalDiagnosis2017}, their effectiveness for disease-symptom extraction remains unclear.

In this paper, we introduce the \collectionlong{} (\collectionshort{}) for the evaluation of graded disease-symptom relations. The collection is annotated by five physicians and contains 235 symptoms for 20 diseases. We label the symptoms using graded judgments~\cite{kekalainen2005binary}, where we differentiate between: \textit{relevant symptoms} (graded as 1) and \textit{primary symptoms} (graded as 2). Primary symptoms---also called cardinal symptoms---are the 
leading symptoms that guide physicians in the process of disease diagnosis. The graded judgments allow us for the first time to measure the importance of different symptoms with grade-based metrics, such as nDCG~\cite{jarvelin2002cumulated}.

As baselines, we implement two methods from previous studies to compute graded disease-symptom relations: In the first method~\cite{shahNeuralNetworksMining2019}, the relation is the cosine similarity between the word vectors of a disease and a symptom, taken from a word embedding model. In the second method~\cite{zhouHumanSymptomsDisease2014}, the relation between a disease and symptom is calculated based on their co-occurrence in the MeSH-keywords\footnote{MeSH-keywords are meta-data that indicates the core topics of an medical article.} of medical articles. We describe limitations of the keyword-based method~\cite{zhouHumanSymptomsDisease2014} and propose an adaption in which we calculate the relations not only on keywords of medical articles, but also on the full text and the title.

We evaluate the baselines on the \collectionshort{} to compare their effectiveness in the extraction of graded disease-symptom relations. As evaluation metrics, we consider precision, recall, and nDCG. For all three metrics, our proposed adapted version of the keyword-based method outperforms the other methods, providing a strong baseline for the \collectionshort{}.

\mbox{}\\\noindent The contributions of this paper are the following:
\begin{itemize}
	\item We introduce the \collectionshort{} for the evaluation of graded disease-symptom relations. We make the collection freely available to the research community.\footnote{Contact this paper's first author to gain access.}
	\item We compare various baselines on the \collectionshort{} to give insights on their effectiveness in the extraction of disease-symptom relations.
\end{itemize}
\vspace{-20pt}
\section{Disease-Symptom Relation Collection}
\vspace{-5pt}
\label{sec:ground_truth}
In this section, we describe the new \collectionlong{} (\collectionshort{}) for the evaluation of disease-symptom relations. We create the collection in two steps: In the first step, relevant disease-symptom pairs (e.g. \textit{appendicitis-nausea}) are collected by two physicians. They collect the pairs in a collaborative effort from high-quality sources, including medical textbooks and an online information service\footnote{The website \textit{netdoktor.at} which is certificated by the Health on the Net Foundation.} that is curated by medical experts.

In the second step, the primary symptoms of the collected disease-symptom pairs are annotated. The annotation of primary symptoms is conducted to incorporate a graded relevance information into the collection. For the annotation procedure, we develop guidelines that briefly describe the task and an online annotation tool. Then, the annotation of primary symptoms is conducted by three physicians. The final label is obtained by a majority voting. Based on the labels obtained from the majority voting, we assign the relevance score 2 to \textit{primary symptoms} and 1 to the other symptoms, which we call \textit{relevant symptoms}.

In total, the \collectionshort{} contains relevant symptoms and primary symptoms for 20 diseases. We give an overview of the collection in Table~\ref{tab:groundtruth_overview}. For the 20 diseases, the collection contains a total of 235 symptoms, of which 55 are labeled as primary symptom (about 25\%). The top-3 most occurring symptoms are: \textit{fatigue} which appears for 15 of the 20 diseases, \textit{fever} which appears for 10, and \textit{coughing} which appears for 7. Notice that the diseases are selected from different medical disciplines: mental (e.g. \textit{Depression}), dental (e.g. \textit{Periodontitis}), digestive (e.g. \textit{Appendicitis}), and respiration (e.g. \textit{Asthma}).

\begin{table}
	\vspace{-15pt}
	\centering
	\caption{Overview of the \collectionshort{}. For each disease, we display the number of relevant symptoms (\#S), the number of primary symptoms (\#P), and the Fleiss' inter-annotator agreement ($\kappa$).}\label{tab:groundtruth_overview}
	\setlength\tabcolsep{5pt}
	\begin{tabular}{lrrr||lrrr}
		\toprule
		\textbf{Disease}                           &   \textbf{\#S} &   \textbf{\#P} &   \boldmath$\kappa$ & \textbf{Disease}              &   \textbf{\#S} &   \textbf{\#P} &   \boldmath$\kappa$ \\
		\hline
		Anorexia Nervosa                   &    7 &    2 & 1.00 & Influenza             &   11 &    2 & 0.57 \\
		Appendicitis                       &    7 &    2 & 1.00 & Measles               &    9 &    4 & 0.38 \\
		Asthma                             &    9 &    4 & 0.76 & Mental Depression     &   13 &    3 & 0.21 \\
		Bronchitis                         &    9 &    1 & 0.71 & Migraine Disorders    &   12 &    4 & 0.37 \\
		Cholecystitis                      &   12 &    1 & 0.55 & Myocardial Infarction &   11 &    4 & 0.44 \\
		COPD &    7 &    3 & 0.83 & Periodontitis         &    3 &    4 & 0.46 \\
		Diabetes Mellitus                  &   11 &    3 & 0.72 & Pulmonary Embolism    &   13 &    2 & 0.83 \\
		Epididymitis                       &    8 &    2 & 0.67 & Sleep Apnea Syndromes &   13 &    2 & 0.31 \\
		Erysipelas                         &    7 &    3 & 0.69 & Tonsillitis           &    7 &    4 & 0.63 \\
		GERD    &    8 &    2 & 0.76 & Trigeminal Neuralgia  &    3 &    3 & 0.28 \\
		\bottomrule
	\end{tabular}
	\vspace{-15pt}
\end{table}

We calculate the inter-annotator agreement using Fleiss' kappa~\cite{fleissMeasuringNominalScale1971}, a statistical measure to compute the agreement for three or more annotators. For the annotation of the primary symptoms, we measure a kappa value of $\kappa=0.61$, which indicates a substantial agreement between the three annotators~\cite{landisMeasurementObserverAgreement1977}. Individual $\kappa$-values per disease are reported in Table~\ref{tab:groundtruth_overview}. By analyzing the disagreements, we found that the annotators labeled primary symptoms with varying frequencies: The first annotator annotated on average 2.1 primary symptoms per disease, the second 2.8, and the third 3.8.

\paragraph{Vocabulary Compatibility:} We map each disease and symptom of the collection to the Unified Medical Language System (UMLS) vocabulary. The UMLS is a compendium of over 100 vocabularies (e.g. ICD-10, MeSH, SNOMED-CT) that are cross-linked with each other. This makes the collection compatible with the UMLS vocabulary and also with each of the over 100 cross-linked vocabularies.

Although the different vocabularies are compatible with the collection, a fair comparison of methods is only possible when the methods utilize the same vocabulary since the vocabulary impacts the evaluation outcome. For instance, the symptom \textit{loss of appetite} is categorized as a symptom in MeSH; whereas, in the cross-linked UMLS vocabulary, it is categorized as a disease. Therefore, the symptom \textit{loss of appetite} can be identified when using the MeSH vocabulary, but it cannot be identified when using the UMLS vocabulary.

\paragraph{Evaluation:} We consider following evaluation metrics for the collection: Recall@k, Precision@k, and nDCG@k at the cutoff $k=5$ and $k=10$. \textit{Recall} measures how many of the relevant symptoms are retrieved, \textit{Precision} measures how many of the retrieved symptoms are relevant, and finally, \textit{nDCG} is a standard metric to evaluate graded relevance~\cite{kekalainen2005binary}.
\vspace{-10pt}
\section{Disease-Symptom Extraction Methods}
\vspace{-5pt}
\label{sec:methods}

\subsection{Related Methods}
\vspace{-5pt}
In this section, we discuss disease-symptom extraction methods used in previous studies. A commonly used resource for the extraction of disease-symptom relations are the articles of the PubMed database. PubMed contains more than 30 million biomedical articles, including the abstract, title, and various meta-data. Previous work~\cite{martinSymptomExtractionIssue2014,hassanExtractingDiseaseSymptomRelationships2015} uses the abstracts of the PubMed articles together with rule-based approaches. In particular, Hassan et al.~\cite{hassanExtractingDiseaseSymptomRelationships2015} derive patterns of disease-symptom relations from dependency graphs, followed by the automatic selection of the best patterns based on proposed selection criteria. Martin et al.~\cite{martinSymptomExtractionIssue2014} generate extraction rules automatically, which are then inspected for their viability by medical experts. Xia et al.~\cite{xiaMiningDiseaseSymptomRelation2018} design special queries that include the name and synonyms of each disease and symptom. They use these queries to return the relevant articles, and use the number of retrieved results to perform a ranking via Pointwise Mutual Information (PMI). 

The mentioned studies use resources that are not publicly available, i.e., rules in~\cite{martinSymptomExtractionIssue2014,hassanExtractingDiseaseSymptomRelationships2015} and special queries in~\cite{xiaMiningDiseaseSymptomRelation2018}. To enable reproducibility in future studies, we define our baselines based on the methods that only utilize publicly available resources, described in the next section.

\vspace{-5pt}
\subsection{Baseline Methods}
\vspace{-5pt}
\label{sec:baselines}
Here, we first describe two recently proposed methods~\cite{shahNeuralNetworksMining2019,zhouHumanSymptomsDisease2014} for the extraction of disease-symptom relations as our baselines. Afterwards, we describe limitations of the method described in~\cite{zhouHumanSymptomsDisease2014} and propose an adapted version in which the limitations are addressed. We apply the methods on the the open-access subset of the PubMed Central (PMC) database, containing 1,542,847 medical articles. To have a common representation for diseases/symptoms across methods (including an unique name and identifier), we consider the 382 symptoms and 4,787 diseases from the Medical Subject Headings (MeSH) vocabulary~\cite{zhouHumanSymptomsDisease2014}. Given the set of diseases ($X$) and symptoms ($S$), each method aims to compute a relation scoring function $\lambda(x,s) \in \mathbb{R}$ between a disease $x \in X$ and a symptom $s \in S$. In the following, we explain each method in detail.

\vspace{-5pt}
\paragraph{$\embedding${\normalfont:}}
Proposed by Shah et al.~\cite{shahNeuralNetworksMining2019}, the method is based on the cosine similarity of the vector representations of a disease and a symptom. We first apply MetaMap~\cite{aronsonEffectiveMappingBiomedical2001}, a tool for the identification of medical concepts within a given text, to the full text of all PMC articles to substitute the identified diseases/symptoms by their unique names. Then, we train a word2vec model~\cite{mikolovDistributedRepresentationsWords2013} with 300 dimensions and a window size of 15, following the parameter setting in~\cite{shahNeuralNetworksMining2019}. Using the word embedding, the disease-symptom relation is defined as $\lambda(x,s)=\textrm{cos}(\mathbf{e}_x, \mathbf{e}_s)$, where $\mathbf{e}$ refers to the vector representation of a word.

\vspace{-5pt}
\paragraph{$\coo${\normalfont:}} This method, proposed by Zhou et al.~\cite{zhouHumanSymptomsDisease2014}, calculates the relation of a disease and a symptom, by measuring the degree of their co-occurrences in the MeSH-keywords of medical articles. The raw co-occurrence of the disease $x$ and symptom $s$, is denoted by $\textrm{co}(x,s)$. The raw co-occurrence does not consider the overall appearance of each symptom across diseases. For instance, symptoms like \textit{pain} or \textit{obesity} tend to co-occur with many diseases, and are therefore less informative. Hence, the raw co-occurrence is normalized by an \textit{Inverse Symptom Frequency (ISF)} measure, defined as $\textrm{ISF}(s)=\frac{|X|}{n_s}$, where $|X|$ is the total number of diseases and $n_s$ is the number of diseases that co-occur with $s$ at least in one of the articles. Finally, the disease-symptom relation is defined as $\lambda(x,s)=\textrm{co}(x,s)\times\textrm{ISF}(s)$. We compute three variants of the $\coo$ method:
\begin{itemize}
    \item $\kwdpmc$: The disease-symptom relations are computed using the MeSH-keywords of the $\approx 1.5$ million PMC articles.
    \item $\kwdpubmed$: While $\kwdpmc$ uses the 1.5 million PMC articles, Zhou et al.~\cite{zhouHumanSymptomsDisease2014} apply the exact same method on the $\approx 30$ million articles of the PubMed database. While they did not evaluate the effectiveness of their disease-symptom relation extraction method, they published their relation scores which we will evaluate in this paper.
    \item $\full$: Applying the $\coo$ method only on MeSH-keywords has two disadvantages: First, keywords are not available for all articles (e.g. only 30\% of the $\approx 1.5$ million PMC articles have keywords) and second, usually only the core topics of an article occur as keywords. We address these limitations by proposing an adaption of the $\coo$ method, in which we use the full text, the title, and the keywords of the $\approx 1.5$ million PMC articles. Specifically, we adapt the computation of the co-occurrence $\textrm{co}(x,s)$, as follows: We first retrieve a set of relevant articles to a disease $x$, where an article is relevant if the disease exists in either the keyword, or the title section of the article. Given these relevant articles and a symptom $s$, we compute the adapted co-occurrence $\textrm{co}(x,s)$, which is the number of relevant articles in that the symptom occurs in the full text. The identification of the diseases in the title and symptoms in the full text is done using the MetaMap tool~\cite{aronsonEffectiveMappingBiomedical2001}.
\end{itemize}{}
\vspace{-16pt}
\section{Evaluation Results \& Discussion}
\vspace{-5pt}
\label{sec:results}
We now compare the disease-symptom extraction baselines on the proposed \collectionshort{}. The results for various evaluation metrics are shown in Table~\ref{tab:results}. The $\full$-variant of the $\coo$ method outperforms the other baselines on all evaluation metrics. This demonstrates the high effectiveness of our proposed adaption to the $\coo$ method.

Further, we see a clear advantage of the $\coo$-method with MeSH-keywords from  $\approx 30$ million PubMed articles as the resource ($\kwdpubmed$) -- in comparison to the same method with keywords from approximately 1.5 million PMC articles ($\kwdpmc$). This highlights the importance of the number of input samples to the method. 
\begin{table}
    \vspace{-23pt}
	\centering
	\caption{Comparison of the disease-symptom extraction methods using our proposed \collectionshort{}. We show significant improvements with: $a$ refers to \small{$\embedding$}, $b$ to \small{$\kwdpmc$}, and $c$ to \small{$\kwdpubmed$} (two-sided, paired t-test: $p<0.01$).}\vspace{0.1cm} \label{tab:results}
	\setlength\tabcolsep{2.1pt}
\begin{tabular}{l|lll|lll}
	\toprule
	\textbf{Method}     &\textbf{nDCG@5}         & \textbf{P@5}           & \textbf{R@5}          & \textbf{nDCG@10}       & \textbf{P@10}         & \textbf{R@10}  \\ \midrule 
	$\embedding$ 		& 0.20                   & 0.18                   & 0.08                  & 0.19                   & 0.15                  & 0.13                \\ 	
	$\copmc$            & 0.27                   & 0.22                   & 0.10                  & 0.22                   & 0.14                  & 0.12                \\ 
	$\copubmed$         & 0.32$^{a}$             & 0.27                   & 0.12                  & 0.28$^{ab}$            & 0.19                  & 0.17        \\ 
	$\cofull$           & \textbf{0.41}$^{abc}$  & \textbf{0.39}$^{abc}$  & \textbf{0.17}$^{abc}$ & \textbf{0.36}$^{abc}$  & \textbf{0.28}$^{abc}$ & \textbf{0.25}$^{ab}$         \\ \bottomrule
\end{tabular}
\vspace{-23pt}
\end{table}

\paragraph{Error Analysis:}
A common error source is a result of the fine granularity of the symptoms in the medical vocabularies. For example, the utilized MeSH vocabulary contains the symptoms \textit{abdominal pain} and \textit{abdomen, acute}\footnote{Symptom for acute abdominal pain}. Both symptoms can be found in the top ranks of the evaluated methods for the disease appendicitis (see Table~\ref{tab:case}). However, since the corpus is not labeled on such a fine-grained level, the symptom \textit{abdomen, acute} is counted as a false positive.

\begin{table}
	\vspace{-23pt}
	\centering
	\caption{Top-4 extracted symptoms of each method for the disease \textit{appendicitis}. The retrieved \colorbox{yellow!30}{relevant symptoms} and \colorbox{green!30}{primary symptoms} are highlighted.} \label{tab:case}
	\setlength\tabcolsep{2.0pt}
	\begin{tabular}{c|c|c|c}
		\toprule
		$\embedding$                        & $\kwdpmc$                           & $\kwdpubmed$                        & $\full$                             \\
		\hline
		Abdomen, Acute                      & \colorbox{green!30}{Abdominal Pain} & Abdomen, Acute                      & Abdomen, Acute                      \\
		\colorbox{green!30}{Abdominal Pain} & Abdomen, Acute                      & \colorbox{green!30}{Abdominal Pain} & \colorbox{green!30}{Abdominal Pain} \\
		Fever of Unknown Origin             & Obesity                             & \colorbox{yellow!30}{Pelvic Pain}   & \colorbox{yellow!30}{Vomiting}      \\
		Renal Colic                         & Thinness                            & Pain, Postoperative                 & \colorbox{yellow!30}{Nausea}        \\
		\bottomrule
	\end{tabular}
	\vspace{-18pt}
\end{table}

Another error source is a result of the bias in medical articles towards specific disease-symptom relationships. For instance, between the symptom \textit{obesity} and \textit{periodontitis}\footnote{A dental disease where the gum that surrounds the teeth retreats} a special relationship exists, which is the topic of various publications. Despite obesity not being a characteristic symptom of a periodontitis, all methods return the symptom in the top-3 ranks. A promising research direction is the selective extraction of symptoms from biomedical literature by also considering the context (e.g. in a sentence) in that a disease/symptom appears.
\vspace{-18pt}
\section{Conclusion}
\vspace{-6pt}
\label{sec:concl}
We introduced the \collectionlong{} (\collectionshort{}) for the evaluation of graded disease-symptom relations. We provided baseline results for two recent methods, one based on word embeddings and the second on the co-occurrence of MeSH-keywords of medical articles. We proposed an adaption to the co-occurrence method to make it applicable to the full text of medical articles and showed significant improvement of effectiveness over the other methods.

\bibliographystyle{splncs04}
\bibliography{bib}

\end{document}